\documentclass[aps,prl,twocolumn,showpacs,preprintnumbers,amsmath,amssymb,a4paper]{revtex4-1}

\begin{document}

\def\pnote#1{{\bf P:#1}}
\def\tr{\textrm{tr}}
\def\cN{\mathcal{N}}
\def\cR{\mathcal{R}}
\def\cF{\mathcal{F}}

%%%%%%%%%%%%%%%%%%%%%%%%%%%%%%%%%%%%%%%%%%%%%%%%%%%%%%%%%%%%%%%%%%%%%%%%%%%%%%%
\preprint{IHES/P/10/13, IPHT-T-/10/045}
\title{The critical ultraviolet behaviour of $\cN=8$ supergravity amplitudes} 
\author{\bf Pierre Vanhove}
\affiliation{  IHES,  Le  Bois-Marie, 35  route de  Chartres,
F-91440 Bures-sur-Yvette, France\\
 Institut de Physique Th{\'e}orique, CEA/Saclay, F-91191 Gif-sur-Yvette,
 France}

\begin{abstract}
  We analyze  the critical ultraviolet behaviour  of the four-graviton
  amplitude in $\cN=8$ supergravity  to all order in perturbation.  We
  use the  Bern-Carrasco-Johansson diagrammatic expansion  for $\cN=8$
  supergravity  multiloop  amplitudes,  where  numerator  factors  are
  squares   of  the   Lorentz  factor   of   $\cN=4$  super-Yang-Mills
  amplitudes, and  the analysis of the  critical ultraviolet behaviour
  of  the   multiloop  four-gluon   amplitudes  in  the   single-  and
  double-trace sectors.   We argue  this implies that  the superficial
  ultraviolet behaviour  of the four-graviton  $\cN=8$ amplitudes from
  four-loop   order   is   determined   by  the   factorization   the
  $\partial^8\,   \cR^4$  operator.   This   leads  to   a  seven-loop
  logarithmic  divergence  in  the  four-graviton  amplitude  in  four
  dimensions.
\end{abstract}

\maketitle

%%%%%%%%%%%%%%%%%%%%%%%%%%%%%%%%%%%%%%%%%%%%%%%%%%%%%%%%%%%%%%%%%%%%%%%%%%%%%
\section{Introduction}
%%%%%%%%%%%%%%%%%%%%%%%%%%%%%%%%%%%%%%%%%%%%%%%%%%%%%%%%%%%%%%%%%%%%%%%%%%%%%

There  have been  recent tremendous  progresses in  the  evaluation of
multiloop    amplitudes    in    maximally    supersymmetric    string
theory~\cite{BerkovitsBT,GreenGT,BerkovitsAW}, in $\cN=8$ supergravity
in  various dimensions~\cite{BernKDD,Bern3L,Bern4L}, and  the analysis
of  the  constraints  from  the  extended  supersymmetry  on  possible
counterterms                       to                      ultraviolet
divergences~\cite{HoweUI,Kallosh:1980fi,BossardMN,Elvang:2010jv}.    It
has  been shown  that up  to and  including four-loop  order  that the
four-graviton  amplitudes in $\cN=8$  supergravity in  four dimensions
are free of ultraviolet divergences~\cite{BernKDD,Bern3L,Bern4L}.  One
important  question  is  to   determine  when  the  first  ultraviolet
divergence appears  in four  dimensions.  In this  work we  argue that
from  four-loop  order  the  critical  ultraviolet  behaviour  of  the
four-graviton  amplitude is  controlled  by the  factorization of  the
$\partial^8\,  \cR^4$  operator.   This  implies that  the  seven-loop
four-graviton  amplitude is  logarithmically ultraviolet  divergent in
four dimensions  with a  counterterm given by  the $\partial^8\,\cR^4$
interaction.

%%%%%%%%%%%%%%%%%%%%%%%%%%%%%%%%%%%%%%%%%%%%%%%%%%%%%%%%%%%%%%%%%%%%%%%%%%%%%
\section{Superficial ultraviolet behaviour of $\cN=8$ amplitudes}
%%%%%%%%%%%%%%%%%%%%%%%%%%%%%%%%%%%%%%%%%%%%%%%%%%%%%%%%%%%%%%%%%%%%%%%%%%%%%

The mass dimension of the $L$-loop gravity amplitude in $D$ dimensions is
given by 
\begin{equation}\label{eDD}
[\mathfrak M_{n,L}^{(D)}]= {\rm mass}^{(D-2)L+2}\,.
\end{equation}
In $\cN=8$ supergravity (and superstring theory) half of the supersymmetry are explicitly
realized  at  each loop  order  and  any  amplitude between  any  four
massless states
$\phi_1$,\dots,$\phi_4$ in the  $\cN=8$ supergraviton multiplet take the form
\begin{equation}\label{eGenericL}
\mathfrak{M}^{(D)}_{4,L}(\phi_1,\dots,\phi_4)={\bf R}^4\, {\cal I}^{(D)}_{4,3}(k^1,\dots,k^4)\,,
\end{equation}
where $ {\cal
I}^{(D)}_{4,L}(k^1,\dots,k^4)$  does not depend  on the  helicities of
the external states, and
is of
superficial  mass dimension  $(D-2)L-6$. The  operator  $ {\bf
  R}^4$  is the  dimension eight  operator defined  from  the massless
four-point $\cN=8$ supergravity tree-level amplitude
\begin{equation}\label{eRfh}
 {\bf R}^4=stu\, \mathfrak M^{(D)}_{4,\textrm tree}(\phi_1,\dots,\phi_4)\,,
\end{equation}
where $s=-(k^1+k^2)^2$, $t=-(k^1+k^4)^2$ and $u=-(k^1+k^3)^2$.
 In the case of the four-graviton
amplitudes the tensorial factor in~(\ref{eRfh}) will be denoted $\cR^4$.
If one parameterises the superficial power counting of the ultraviolet behaviour of the amplitude as
\begin{equation}\label{eAmpD}
[\mathfrak{M}_{4,L}^{(D)}]= \Lambda^{(D-2)L-6-2\beta_L}\, \partial^{2\beta_L} \cR^4\,,
\end{equation}
where $\Lambda$ is a momentum cutoff, 
the   critical   dimension   for   ultraviolet  divergences   in   the
four-graviton amplitude is given by
\begin{equation}\label{eDc}
D\geq 2+{6+2\beta_L\over L}\,.
\end{equation}
Up to  an including four-loop  the supersymmetry constraints~\cite{BerkovitsBT,GreenGT}  
implies that $\beta_L=L$. 

\nobreak
If  only  the large-$\lambda$  regulator  of  the  pure spinor  string
formalism is used one can show~\cite{BerkovitsBT,GreenGT} that $\beta_L=L$ for $L\leq 6$ and
$\beta_L=6$ for  $L\geq 6$ leading  to the critical dimension  for the
appearance ultraviolet
divergence
\begin{equation}\label{eDcSix}\begin{split}
D\geq 4+{6\over L}&; \qquad {\rm ~for~} L\leq 6\cr
D\geq 2+{18\over L}&; \qquad {\rm ~for~} L\geq 6\,.\cr
\end{split}\end{equation}
This rule implies a logarithmic nine-loop divergence in the four-graviton
amplitude in four dimensions~\cite{GreenGT}.
From genus five  possible divergences from the tip  of the pure spinor
cone that would require the use of the complicated small-$\lambda$ 
regulator~\cite{BerkovitsVI,GrassiFE}  can  restrict  $\beta_L=4$  for
$L\geq4$ leading to the critical dimension for the appearance of ultraviolet
divergence
\begin{equation}\label{eDcFour}\begin{split}
D\geq 4+{6\over L}&; \qquad {\rm ~for~} L\leq 4\cr
D\geq 2+{14\over L}&; \qquad {\rm ~for~} L\geq 4\, .\cr
\end{split}\end{equation}
This rule implies a logarithmic seven-loop divergence in the four-graviton
amplitude in four dimensions.

The  issue of correctly  identifying the  ultraviolet behaviour  of a
supersymmetric  theory is  equivalent  to the  understanding of  which
interactions are true F-terms satisfying non-renormalisation theorems, and
which  interactions  are D-terms  receiving  quantum  corrections to  all
orders  in  perturbation~\cite{Grisaru:1979wc}.  We  now  discuss  the
D-terms   arising   in  the   four   points   amplitudes  in   $\cN=4$
super-Yang-Mills (SYM) in various dimensions before turning to the case of  $\cN=8$ supergravity.

%%%%%%%%%%%%%%%%%%%%%%%%%%%%%%%%%%%%%%%%%%%%%%%%%%%%%%%%%%%%%%%%%%%%%%%%%%%%%
\section{Parameterization of the $\cN=4$ super-Yang-Mills amplitudes} 
%%%%%%%%%%%%%%%%%%%%%%%%%%%%%%%%%%%%%%%%%%%%%%%%%%%%%%%%%%%%%%%%%%%%%%%%%%%%%

 By   partial  integration   over  the
superspace  variables it  is possible  to rewrite  D-term  as ``fake''
F-term, and  detecting the  true D-term nature  of an interaction  can be
non-trivial.  In the four-point open string amplitudes with $U(N)$ gauge group
the   true  D-term  nature   of  the   $\partial^2t_8\tr  F^4   $  and
$\partial^4t_8(\tr F^2)^2$ interactions
became manifest by the appearance of inverse
derivative factors arising when integrating over the string theory
moduli~\cite{BerkovitsAW}. It was shown in this analysis that
at three-loop  order the  integration over
the  open  string  moduli  generates  one  inverse  derivative  factor
$1/(k^a\cdot k^b)$
in  the single  trace sector.  From four-loop  two  inverse derivative
factors $1/(k^a\cdot k^b)\times 1/(k^c\cdot k^d)$ arise in the single trace and the double-trace sectors. It is important to realize
that these inverse derivative contributions do not lead to poles in the total
amplitude, they only reduce  the number of external momenta factorizing
the amplitude.

One can represent  the four-gluon amplitudes in $\cN=4$  SYM from four-loop order  in a form where  the ultraviolet behaviour  is explicit in
single-trace and double-trace sectors
\begin{widetext}\begin{equation}\begin{split}
  \label{e:AmpTrace}
  g_{\rm          YM}^{-2L-2}\,\mathfrak{A}_{4,L}^{(D)}&=t_8(F^1F^2F^3F^4) \tr(T^1T^2T^3T^4)\,(k^1\cdot k^2)\,\int\prod_{i=1}^L
    {d^D\ell_i\over(2\pi)^D}\,{t_1(\ell_i,k^i)\over
      \prod_{j=1}^{3L+1}\ell_{j}^2}+perm(2,3,4)\cr
&+t_8(F^1F^2F^3F^4) \,\tr (T^1T^2)\tr(T^3T^4)\,(k^1\cdot k^2)\, k^a_mk^b_n\,\int\prod_{i=1}^L
    {d^D\ell_i\over(2\pi)^D}\,{t_2^{mn|ab}(\ell_i,k^i)\over
      \prod_{j=1}^{3L+1}\ell_{j}^2}+perm(2,3,4)\,.
\end{split}\end{equation}
\end{widetext}
The Lorentz scalar $\cF^4\equiv t_8(F^1F^2F^3F^4)$ is defined by the tree-level four-gluon amplitude 
\begin{equation}\label{e:treeYM}
 \mathfrak A^{(D)}_{4,\textrm{tree}}= g_{\rm YM}^2\,{\cF^4\over stu}\, [u\,
\tr(T^1T^2T^3T^4)
 +perm(2,3,4)]\,.
\end{equation}
For large  values of the  loop momenta $\ell_i$ the  numerator factors
in~(\ref{e:AmpTrace}) behave as  $t_1(\ell_i,k^i)\sim\ell^{2L-4}$  and  $t_2^{mn|ab}(\ell_i,k^i)\sim\ell^{2L-6}$   . [We have use made explicit the presence of a
$s=-2(k^1\cdot      k^2)$      factor      in     the      $s$-channel
diagram~\cite{BernKDD,Bern3L,Bern4L}.  The    $t$- and  $u$-channel
have the same property after permutation of the labels of the external states.]
 The $\partial^2\,t_8\tr F^4$ carries the critical ultraviolet behaviour of the
 four-point amplitudes from two-loop order that have ultraviolet
 divergences in $D\geq 4+6/L$ dimensions.
The $\partial^4\,t_8(\tr F^2)^2$ carries the critical ultraviolet behaviour for the
  double-trace   contribution  of   the   four-point  amplitude   from
  three-loop order which has ultraviolet divergences in $D\geq 4+8/L$ dimensions~\cite{BerkovitsAW}.

  A different parameterization of  Yang-Mills amplitudes based on cubic
  interactions have been introduced by Bern, Carrasco and Johansson at
  tree-level      in~\cite{Bern:2008qj}       and    at  loop      orders
  in~\cite{Bern:2010ue,Bern:2010yg}.   In  this   parameterization  the
  $L$-loop four-point amplitudes take the form
\begin{equation}
  \label{e:SYMbcj}
  \mathfrak    A_{4,L}^{(D)}=   g_{\rm   YM}^{2L+2}\,    \sum_j   \int
  \prod_{i=1}^L        {d^D\ell_i\over(2\pi)^D}        \,        {1\over
    S_j}\,{n_jc_j\over\prod_{r=1}^{3L+1} p^2_r}\,,
\end{equation}
where the sums now run  over all distinct $n$-point $L$-loop diagrams with
cubic vertices. The inverse propagators of the diagram are $p_r^2$, $c_i$ are color factors, $n_i$ are Lorentz factors and $S_j$ are
symmetry factors of the corresponding Feynman diagrams~\cite{Bern:2010ue}.  The
Lorentz factors $n_i$ are constrained by the Jacobi-like relations
$n_i+n_j+n_k=0$  similar to  the one  satisfied by  the  color factors
$c_i+c_j+c_k=0$.  Such relations, that are necessary for a correct
parameterization the  amplitudes, where  shown to be  a consequence  of the
monodromy         relations         between         open         string
amplitudes~\cite{BjerrumBohr:2009rd} (see as well~\cite{Stieberger:2009hq,Mafra:2009bz,Tye:2010dd}).

Instead of the form of the amplitude given in~(\ref{e:AmpTrace}), where
the ultraviolet behaviour is explicit, we express the higher-loop SYM
amplitudes  on a  basis  of integrals  functions containing  1-particle
reducible graphs with $1/(k^a\cdot k^b)$ factors cancelling 
corresponding contributions in the numerators. This way we are mimicking the
appearance of the inverse derivative factors noticed in the open string analysis~\cite{BerkovitsAW}.

 In this basis a first class of integral functions contributing
 to the sum in~(\ref{e:SYMbcj}), is composed by 
 1-particle irreducible (1PI) diagrams with Lorentz factor given by
\begin{equation}
\label{e:n1} n_j^{1PI}= \cF^4\times (k^1\cdot k^2)\, k^a_m k^b_n\,t_{j,1PI}^{mn|ab}(\ell_i,k^i)\,,
\end{equation}
in the $s$-channel.  For the large values of the loop
momenta $\ell_i$  this  factor has the behaviour
\begin{equation}
  \label{e:t1PI}
\lim_{\ell_i\to\infty}  t_{j,1PI}^{mn|ab}(\ell_i,k^i)\sim \ell^{2L-6}\,.
\end{equation}
In order to reproduce the correct ultraviolet behaviour of the $\cN=4$
SYM four-point amplitude~(\ref{e:AmpTrace}), one needs a second class of
integral functions with 1-particle reducible (1PR) graphs with propagator
contributions  of the type $1/(k^a\cdot k^b)$.   As   in  the  open   string  theory  analysis
 of~\cite{BerkovitsAW} these  reducible graphs do not  lead to massless
 pole in the total amplitude because the $1/(k^a\cdot k^b)$ factor are cancelled
 by corresponding factors in the numerator.

At three-loop  order one only  needs  1-particle  reducible integral
functions with  one factor of inverse derivative~\cite{BerkovitsAW,Bern:2010ue}  with numerators given
by the Lorentz factor
\begin{equation}
  \label{e:n3loop}
  n_j^{1PR}|_{L=3}=  \cF^4\times(k^1\cdot k^2)^2\, t^{1PR,L=3}_j(\ell_i,k^i)\,,
\end{equation}
in the $s$-channel. Examples of
such three-loop contributions  are given by the diagrams~$(j)$--$(l)$ of fig.~2
 in~\cite{Bern:2010ue}. For large
values of the loop momenta $\ell_i$ this  factor has the behaviour
\begin{equation}
  \label{e:t1PR3L}
\lim_{\ell_i\to\infty} t^{1PR,L=3}_j(\ell_i,k^i)\sim 1\,.
\end{equation}
  From four-loop order one needs 1-particle reducible integral
 functions with two inverse derivative factors~\cite{BerkovitsAW}. 
 Such integral functions  have numerator Lorentz factors
 given by 
\begin{equation}
\label{e:n2} n_j^{1PR}= \cF^4\times(k^1\cdot k^2)^3\, t^{1PR}_j(\ell_i,k^i)\,,
\end{equation}
in the $s$-channel. For large
values of the loop momenta $\ell_i$ this  factor has the behaviour
\begin{equation}
  \label{e:t1PR}
\lim_{\ell_i\to\infty}t^{1PR}_j(\ell_i,k^i)\sim \ell^{2L-8}\, .  
\end{equation}

This parameterization of the four-point multiloop $\cN=4$ SYM amplitude
reproduces the critical ultraviolet behaviour~\cite{BerkovitsAW}.

%%%%%%%%%%%%%%%%%%%%%%%%%%%%%%%%%%%%%%%%%%%%%%%%%%%%%%%%%%%%%%%%%%%%%%%%%%%%%
\section{Parameterization of the $\cN=8$ supergravity amplitudes}\label{sec:Dtermsugra}
%%%%%%%%%%%%%%%%%%%%%%%%%%%%%%%%%%%%%%%%%%%%%%%%%%%%%%%%%%%%%%%%%%%%%%%%%%%%%

The question of D-term and F-term interactions in $\cN=8$ supergravity
can be approached in a similar fashion as in the SYM case discussed in
the     previous    section.      The     pure    spinor     formalism
indicates~\cite{BerkovitsBT}  that the $\cR^4$,  $\partial^4\cR^4$ and
$\partial^6\cR^4$  interactions are  F-terms represented  by integrals
over the pure  spinor superspace involving only the  regulator for the
large-value  of   the  pure  spinor   ghost~\cite{BerkovitsBT}.  These
interactions  receive only  a  finite number  of  loop corrections  in
string  perturbation~\cite{GreenBF,GreenGT}.   There  are  indications
from string/M-theory  computations that $\partial^8\,\cR^4$  term is a
D-terms~\cite{GreenBF}.  The analysis  using the pure spinor formalism
indicates   that   from  five-loop   order   the  amplitudes   receive
contributions   from  the   small-$\lambda$   regulator  and   inverse
derivative   factors   from    the   integration   over   the   string
moduli~\cite{BerkovitsAW}.

With the parameterization~(\ref{e:SYMbcj}) of the Yang-Mills amplitudes
it was  proposed in~\cite{Bern:2010ue} that  the multiloop
four-point supergravity amplitudes take the symmetric form
\begin{equation}
  \label{e:SUGRAbcj}
  \mathfrak    M_{4,L}^{(D)}=   \kappa_{(D)}^{2L+2}\,    \sum_j   \int
  \prod_{i=1}^L        {d^D\ell_i\over(2\pi)^D}        \,        {1\over
    S_j}\,{n_j\tilde n_j\over\prod_{r=1}^{3L+1} p^2_{r}}\,.
\end{equation}
This  parameterization  of  the  supergravity multiloop  amplitudes  is
particularly  well  suited for  incorporating  the inverse  derivative
effects.  The left-moving $n_j$  and right-moving $\tilde n_j$ Lorentz
factors have a
generalized      gauge     freedom     leaving      the     amplitudes
invariant~\cite{Bern:2010ue,Bern:2010yg}. We make the choice 
 that $\tilde n_j$ is $n_j$ expressed in terms of the right
moving polarisations.

For  the   1-particle  irreducible  integral   functions  contribution
to~(\ref{e:SUGRAbcj}) the numerators are given by the square of the SYM
Lorentz factors in~(\ref{e:n1})
 \begin{equation}
   \label{e:SUGRAnum1}
   n_i^{1PI}\tilde n_i^{1PI}\sim k^8 \, \cR^4\times (t^{mn|ab}_{j,1PI}(\ell_i,k^i))^2\,.
 \end{equation}
 Where  we  used that  $\cR^4=\cF^4\tilde  \cF^4$  which  is a  direct
 consequence  of  the  KLT  relation~\cite{Kawai:1985xq}  between  the
 four-point  tree-level  amplitude   in  gravity~(\ref{eRfh})  and  in
 Yang-Mills~(\ref{e:treeYM}).  Using  the behaviour~(\ref{e:t1PI}) for
 the large values of the loop momenta $\ell_i\sim \Lambda\gg 1$, the
 superficial  ultraviolet  behaviour  of  the  1-particle  irreducible
 integral          function          in~(\ref{e:SUGRAbcj})          is
 $\partial^8\,\cR^4\,\Lambda^{(D-2)L-14}$.

The three-loop supergravity amplitude~(\ref{e:SUGRAbcj}) contains the 1-particle reducible graphs bringing
$1/(k^a\cdot k^b)$ contributions~\cite{BerkovitsAW,Bern:2010ue}.  The numerators
are given by the square of the SYM Lorentz factors in~(\ref{e:n3loop})
 \begin{equation}
   \label{e:SUGRAnum2}
   n_j^{1PR}|_{L=3}\tilde n_j^{1PR}|_{L=3}\sim k^{8} \, \cR^4\times(t_j^{1PR,L=3}(\ell_i,k^i))^2\,.
 \end{equation}
Because      these  1-particle
reducible graphs have a single $1/(k^a\cdot k^b)$ factor and using the
behaviour~(\ref{e:t1PR3L}) for
 the large values of the loop momenta $\ell_i\sim \Lambda\gg 1$, the superficial ultraviolet behaviour of
these diagrams is $\partial^6\,\cR^4\,\Lambda^{3(D-6)}$. This is
the correct leading ultraviolet behaviour, given by~(\ref{eAmpD}) with
$\beta_3=3$,       for        the       three-loop       four-graviton
amplitude~\cite{GreenGT,Bern3L}.
For  $L=3$ the  contribution  from the  1-particle irreducible  graphs
in~(\ref{e:SUGRAnum1})   give   the   milder   ultraviolet   behaviour
$\partial^8\,\cR^4\, \Lambda^{3D-20}$  of the diagrams~$(a)$--$(i)$ in
fig.~2 of~\cite{Bern:2010ue} (see as well~\cite{Bern3L}).

From four-loop order the contributions to the 1-particle reducible
integral functions  to~(\ref{e:SUGRAbcj}) have two  inverse derivative
factor~\cite{BerkovitsAW} with  numerators given by the  square of the
SYM Lorentz factor in~(\ref{e:n2})
 \begin{equation}
   \label{e:SUGRAnum3}
   n_j^{1PR}\tilde n_j^{1PR}\sim k^{12} \, \cR^4\times (t^{1PR}_j(\ell_i,k^i))^2\,.
 \end{equation}
 Because  these  1-particle   reducible  graphs  have  double  inverse
 derivative factors, $1/(k^a\cdot k^b)\times1/(k^c\cdot k^d)$, and using the
 behaviour~(\ref{e:t1PR})  for the  large values  of the  loop momenta
 $\ell_i\sim \Lambda\gg  1$, the superficial  ultraviolet behaviour of
 the supergravity  amplitudes~(\ref{e:SUGRAbcj}) at $L\geq4$  loops is
 $\partial^8\,\cR^4\,\Lambda^{(D-2)L-14}$.  This   correspond  to  the
 critical  ultraviolet behaviour with  $\beta_L=4$ for  $L\geq4$ given
 in~(\ref{eDcFour}).

 Since  the $\cN=8$ supergravity amplitudes in~(\ref{e:SUGRAbcj}) 
 are composed  by integrals with  numerators given  by squares   summed
 over with  positive relative coefficients given  the symmetry factors
 of  the diagrams,  it is  unlikely  that cancellations  leading to  a
 better  ultraviolet   behaviour  can  arise   without  modifying  the
 ultraviolet behaviour  of the  $\cN=4$ SYM amplitudes.   We therefore
 conclude   that  the   superficial  ultraviolet   behaviour   of  the
 four-graviton $\cN=8$  supergravity amplitudes  is given by  the rule
 in~(\ref{eDcFour}).   This implies  that the  four-graviton amplitude
 has a  seven-loop logarithmic divergence in four  dimensions with the
 $\partial^8\,\cR^4$   interaction  for   counterterm.    A  candidate
 on-shell superspace $E_7$-invariant expression for this seven-loop counterterm
 is the volume of superspace
\begin{equation}\label{eCT}
\int d^4x \sqrt{-g^{(4)}}\,\partial^8\cR^4 \sim \int d^4x\int
d^{32}\theta \, |E|\,,
\end{equation}
where  $|E|$ is  the determinant  of the  super-vierbein.  To the
contrary to the  volumes of $\cN=1$ and $\cN=2$  superspace this expression
does not seem to be vanishing~\cite{BossardMN}.

\section*{Acknowledgements} 

We would like to thank Henrik Johansson for discussions, and Michael
B.  Green for comments on the draft who informed us   about the  work~\cite{GreenToAppear} where similar
ideas are considered. The author would like to thank the
Laboratoire Poncelet in Moscow for hospitality when 
the paper has been written.

\end{document}